\documentstyle[aps,eqsecnum,multicol,epsf]{revtex}
\begin{document}

\draft
\title{
Line Junctions in the Quantum Hall Effect}

\author{C.L. Kane}
 
\address{Department of Physics, University of Pennsylvania\\
Philadelphia, Pennsylvania 19104
}

\author{Matthew P.A. Fisher}
\address{Institute for Theoretical Physics, University of California,
Santa Barbara, CA 93106--4030}

\date{\today}
\maketitle

\widetext

\begin{abstract}
A long skinny gate across a fractional 
quantum Hall fluid at filling $\nu=1/m$ with odd integer $m$,
creates a novel one-dimensional (1d) system which is
isomorphic to a disordered 1d electron gas with {\it attractive} interactions.
By varying the gate potential along such a line junction,
it should be possible to tune through
the 1d localization transition, predicted for an attractively
interacting electron gas. 
The key signature of this 1d metal-insulator transition
is the temperature dependence of the conductivity,
which diverges as a power of temperature in the metallic phase, and vanishes
rapidly in the insulator.
We show that  
the 1d conductivity can be extracted
from a standard Hall transport measurement, in the regime where the
Hall conductance is close to its quantized value.
A line junction in a $\nu=2/3$ quantized Hall fluid
is predicted to exhibit a similar localization
transition.
\end{abstract}

\begin{multicols}{2}
\narrowtext

\section{Introduction}

Edge states in the quantum Hall effect offer
a highly controlled laboratory for the experimental study of
quantum transport in one dimension.
The right and left moving edge modes, which reside on the
opposite edges of a quantum Hall bar form an ideal one
dimensional electron gas.
Since the edges are spatially
separated from one other, backscattering due to impurities, which
usually localizes electrons in one dimension, may effectively
be eliminated.  

Following Wen's suggestion that the edge states in the fractional quantum
Hall effect are chiral Luttinger liquids\cite{Wen}, there has been considerable
interest in the experimental implications of Luttinger liquid theory on
edge state transport.
Much of the focus has been on the nature of point
contact tunneling.  Specifically, pinching a quantum Hall bar at a 
point using a patterned gate electrode, introduces local and controllable
backscattering between oppositely moving edge modes.
This is analogous to a single impurity in
an otherwise clean one dimensional electron gas\cite{KF1}.  Luttinger liquid
theory predicts that the tunneling conductance through the point contact
vanishes 
as a power of temperature with a universal exponent, which depends on the structure
of the bulk quantum Hall fluid.  Milliken, Umbach and Webb have observed a temperature
dependence consistent with the predicted $T^4$
behavior for tunneling between two $\nu = 1/3$ fluids\cite{Webb}.  More recently,
Chang et. al. have measured the tunneling conductance
between a Fermi liquid and a $\nu = 1/3$ edge
state, and found behavior consistent with the predicted $T^2$ 
temperature dependence\cite{Chang}.

A different and perhaps more interesting way of introducing
intermode backscattering is depicted schematically in  
Figure 1.  A bulk quantum Hall fluid
is divided into two pieces by depleting
the electron gas along a narrow line, using a long ``skinny" gate.
Such a ``line junction", creates ``internal" edge states
which propagate in opposite directions on either side of the
gate.  Together, these two modes constitute a novel (non-chiral)
one-dimensional system - a ``quantum anti-wire". 
As the gate potential is varied, the degree of backscattering between
the two counter-propagating modes can be varied.
For strong depletion under the gate, all backscattering
can be effectively eliminated, and the source to drain conductance
vanishes.  In the opposite limit, the gate potential can be turned off,
and the (two-terminal) source-to-drain
conductance is quantized.  But what happens in between?
For intermediate values of $V_G$, intermode backscattering 
will be mediated by inhomogeneities, either of the gate itself
or due to nearby impurities in the electron gas.

\begin{figure}
\epsfxsize=3in
\centerline{\epsffile{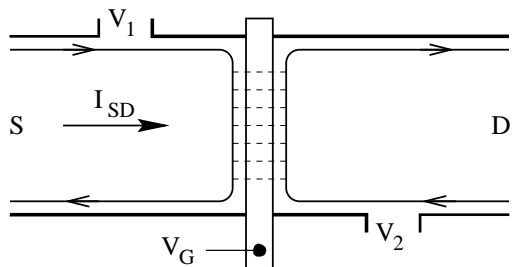}}
\caption{A long skinny gate across a quantum Hall bar creates a line junction, with
oppositely moving edge modes (lines with arrows) on either
side of the gate.  The intermode
backscattering rate (dotted lines) can be varied
by changing the gate potential $V_G$,
which drives a one-dimensional metal-insulator transition.
}
\end{figure}

Since the line junction is effectively a disordered one-dimensional
electron system, one might expect that electron localization is inevitable.
For the integer quantum Hall effect, this expectation is valid.
However, 
Renn and Arovas\cite{Renn} have recently shown that for a
fractional quantum Hall fluid at filling $\nu=1/3$,
the ``anti-wire" line junction is formally equivalent to
a 1d electron gas with {\it attractive} electron interactions.
As shown some years back by Giamarchi and Schulz\cite{Giamarchi}, 
a disordered 1d electron gas becomes
metallic for sufficiently strong attractive interactions - 
that is, all states are {\it not} localized in 1d.
Upon varying the strength of the attractive interaction, a disorder
driven metal-insulator transition was predicted.
This metal-insulator transition should be directly observable in such
a fractional quantum Hall effect line
contact.

In this paper
we describe in detail the experimental signature of a
1d metal-insulator transition for a quantum Hall line junction.  
The transition is conveniently 
characterized by the temperature dependence of a one-dimensional
{\it conductivity}, $\sigma$ - an intensive quantity.
For an infinitely long system,   
Giamarchi and Schulz\cite{Giamarchi} argue that the conductivity vanishes
at $T =0$ in the insulating phase, but diverges  
as $T \rightarrow 0$ in the metallic phase.
However, the most accessible experimental quantity is the source-to-drain
{\it conductance}, for a Hall bar with a finite width, $L$.
Nevertheless, by tuning the gate potential into a regime
where the Hall conductance is close to it's quantized value,
it is possible to extract the ``anti-wire" conductivity,
as we discuss in detail below.

We begin in section II with a review of the Luttinger liquid model
for a line junction and 
show, following Renn and Arovas, that for FQHE
states in the Laughlin sequence, $\nu=1/m$ with odd $m$, a 
1d metal-insulator transition should be accessible.
We describe the temperature dependence of the conductivity in the
metal and insulating phases as well as near the transition,
in section III.  
In section IV we show how the conductivity can be extracted from a Hall conductance measurement.
Finally, in section V we consider the line contact for 
a hierarchical FQHE state at filling $\nu=2/3$, and argue that a similar
metal-insulator transition should occur there as well.

\section{Model and Transition}
The bosonized Hamiltonian density for a clean line junction can
be written in terms of right and left moving electron densities, $n_{R/L}$:
\begin{equation}
{\cal H}_0 = {{\pi v_0} \over \nu} (n_R^2 + n_L^2 + 2\lambda n_R n_L) .
\end{equation}
These densities
satisfy Kac-Moody commutation relations\cite{Wen}:
\begin{equation}
[n_{R/L}(x),n_{R/L}(x^\prime)] = \pm (i\nu /2 \pi ) 
\partial_x \delta(x-x^\prime) .
\end{equation}
When $\lambda =0$ this Hamiltonian describes de-coupled right and left
moving modes, which propagate at a velocity $v_0$.
The term proportional to $\lambda$ represents a screened Coulomb
interaction between the right and left moving modes.
We have assumed that the long-ranged piece
of the Coulomb interaction is screened by a ground plane, or
the line-junction gate itself.

When the gate potential is large, the two modes are well separated
spatially, and the interaction $\lambda$ is small.  As the
gate potential is decreased, the modes move closer together
increasing the repulsive interaction $\lambda$.  But in addition,
tunneling of electrons between
the right and left modes {\it under} the gate becomes possible.
To incorporate these processes we add an additional term to the
Hamiltonian:
\begin{equation}
{\cal H}_1 = \xi(x) \psi_R^\dagger (x) \psi_L(x) + h.c. ,
\end{equation}
where $\psi_R$ is an electron destruction operator in the right
moving mode.  These operators
can be re-expressed in terms of boson fields
$\phi_{R/L}$, which are 
proportional to the electron densities:
\begin{equation}
n_{R/L} = \pm {1 \over {2\pi}} \partial_x \phi_{R/L} .
\end{equation}
Specifically, 
\begin{equation}
\psi_R \sim e^{i \phi_R /\nu} ,
\end{equation} 
and similarly for the left moving mode. 

The electron tunneling amplitude $\xi(x)$ is generally complex.
For a perfectly clean line junction one expects
$\xi(x) \sim e^{i\delta k x}$ where $\delta k$ is  
a gauge invariant momentum difference between the right and left moving modes.
If the edge modes are separated by a distance $d$ then
$\delta k = 2\pi Bd/\Phi_0$, where $B$ is the applied magnetic
field, and $\Phi_0 =hc/e$ is the magnetic flux quantum.
However, in any real device one expects the presence of impurities
near the line junction, which will effect also the magnitude
of the tunneling strength, $|\xi|$.
We thus assume that $\xi(x)$ is a random complex
variable, uncorrelated on length scales
long compared to the inter-impurity spacing ``$a$".  In practice
one expects $a$ to be comparable to or smaller than the distance
to the line-junction gate.   For further simplicity,
we take $\xi(x)$ to have a Gaussian distribution,
\begin{equation}
[\xi(x) \xi^*(x')]_{\rm ens} = \Delta_W \delta(x-x')  ,
\end{equation}
where the square brackets denote an ensemble average over impurity
configurations.  For later convenience we define
a dimensionless impurity strength, $W$
\begin{equation}
W = {a \over \omega_c^2} \Delta_W  .
\end{equation}
Here the cutoff frequency $\omega_c$ is set by the bulk quantum Hall
gap - the cyclotron frequency when $\nu =1$.
Upon increasing the gate potential, which brings the edge modes
closer together enabling tunneling, one expects that the effective disorder
strength, $W$, increases in magnitude.

As emphasized by Renn and Arovas\cite{Renn} the above model for a line junction
is mathematically equivalent to a model of
a one-dimensional interacting
electron gas with impurity scattering present.
For the IQHE at $\nu =1$ the electron gas is repulsively interacting,
and one anticipates that the line junction will be insulating
with all states localized.  But most remarkably, 
when $\lambda$ is small the electron gas isomorphic to
the
FQHE line-junction has {\it attractive} interactions.
To see this we define new non-chiral boson fields,
\begin{equation}
\phi_{R/L} = \sqrt{\pi} (\phi \pm \nu \theta ) ,
\end{equation}
which are canonically conjugate variables,
\begin{equation}
[\theta(x), \partial_x \phi(x')] = i \delta(x-x') .
\end{equation}
In terms of these field the pure Hamiltonian becomes,
\begin{equation}
{\cal H}_0 = {v \over 2} [K (\partial_x \phi)^2 + {1 \over K} (\partial_x
\theta)^2 ] ,
\end{equation}
with a renormalized velocity
\begin{equation}
v = v_0 (1 + \lambda^2)^{1/2} ,
\end{equation}
and a dimensionless ``stiffness" 
\begin{equation}
K = {1 \over \nu} [{{1- \lambda} \over {1+ \lambda}}]^{1/2} .
\end{equation}
The random piece of the Hamiltonian involves only the field $\theta$:
\begin{equation}
{\cal H}_1 = \xi(x) e^{i2\sqrt{\pi}\theta(x)} + h.c. .
\end{equation}
The model is equivalent to a bosonized representation
of an interacting Luttinger liquid with impurity scattering.
The stiffness $K$ is equal to the dimensionless
conductance $g$ for the Luttinger liquid.
Thus, $K<1$ describes a repulsively interacting electron gas,
whereas $K>1$ an attractively interacting gas.

Remarkably, for a $\nu=1/3$ line-junction with well
separated modes (small $\lambda$), 
the equivalent electron gas is strongly attractive, $K=1/\nu$.
This should be contrasted to a very narrow
quantum Hall bar which also has right and left
moving modes.  In this case, the 
dominant inter-mode tunneling process is a fractionally
charged Laughlin quasiparticle.  The system is isomorphic
to a repulsively interacting electron gas with
$K = \nu$, rather than $K=1/\nu$ as above.

The effects of impurity scattering on an interacting Luttinger liquid
has been considered by a number of authors\cite{Giamarchi,Schulzreview,Apel,KF2}.
The renormalization group calculation by Giamarchi and Schulz\cite{Giamarchi}
reveals clearly the phase boundary separating an insulating
from a conducting phase. 
Working in momentum space, they integrate over the
field $\theta(k, \omega_n)$, for
a shell of modes with $\Lambda/b < k < \Lambda$,
and rescale as $k' =bk$ and $\omega_n' = b^z \omega_n$.
Here $\Lambda \sim 1/a$ is a cut-off and $\omega_n$ is
a Matsubara frequency.  The dynamical exponent $z$ is chosen to
keep the velocity $v$ invariant.  To leading order in $W$
the RG recursion relations are ($\ell = {\rm \ln} b$):  
\begin{equation}
\partial W/\partial \ell = (3-2K)W  ,
\end{equation}
\begin{equation}
\partial K/ \partial \ell = -{{K^2} \over 2} W  .
\end{equation}
with $z=1-(KW/2)$.  These equations describe a phase transition
between a conducting phase, in which the disorder strength
$W$ scales to zero, and an insulating disorder dominated phase,
as sketched in Fig. 2.
For small $W$ the phase boundary is at $K=3/2$, and increases
to larger $K$ with increasing $W$.  

For the IQHE line-junction ($\nu =1$), the largest value of $K$ is
one, so that the system is always in the localized phase.
However, for a $\nu=1/3$ FQHE line junction, the maximum value
of $K$ is 3, which occurs when the
modes are well separated and $W$ is small.
This puts the system well into the conducting phase.
With increasing gate potential, both the tunneling
($W$) and the interactions ($\lambda$) increase,
which moves the system along the trajectory sketched in Fig. 2.
The system will undergo a phase transition into a localized
state.   
\begin{figure}
\epsfxsize=3in
\centerline{\epsffile{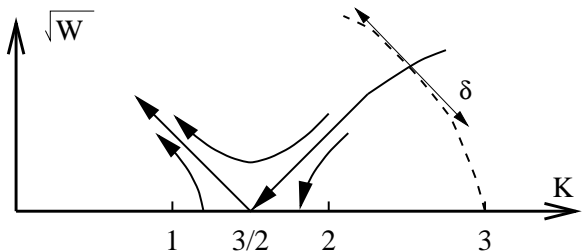}}
\caption{Renormalization Group flow diagram for 1-d metal-insulator transition,
with disorder strength $W$, and interaction parameter $K$.  
The dashed line represents the initial values of $W$ and $K$
for $\nu=1/3$
as the voltage on the line junction gate is varied.
The parameter $\delta$ measures the ``distance" to the transition.}
\end{figure}

This localization transition should be observable in FQHE line junctions.
In the next Section we consider the behavior of the
transport along the line junction,
first under the assumption that
the line junction is infinitely long.  We then
describe the predicted behavior for a finite length
line-junction fed by QHE edge states, as depicted in Fig. 1.
In Section IV we argue that a $\nu=2/3$ line junction should
exhibit a similar localization transition.

\section{Bulk Conductivity}

Transport along the line junction is characterized 
by a one-dimensional conductivity $\sigma$. 
Of interest is the temperature dependence 
in the insulating and conducting phases, as well
as near the transition.
 
In the insulating phase
at low temperatures, the transport presumably takes
place via variable-range hopping processes
between nearby localized states.
This gives
\begin{equation}
\sigma(T) \sim e^{-(T_0/T)^{1/2}} .
\end{equation}
The temperature scale $T_0$ is set by the localization length,
$\xi_{loc}$, varying as $T_0 \sim v/\xi_{loc}$.
Deep within the localized phase,
$\xi_{loc} \sim a$ and the temperature
scale should be large.
Upon approaching the transition from the insulating side,
the localization length diverges as 
\begin{equation}
\xi_{loc} \sim a e^{c/\delta^{1/2}}  ,
\end{equation}
where $\delta$ is the distance to the transition and $c$ is a constant.
The parameter $\delta$ may be tuned by varying the gate voltage $V_G$, $\delta \propto V_G - V_{Gc}$.

In the conducting phase, the disorder $W$ scales to zero, and
the conductivity should be infinite at zero temperature.
Finite temperature cuts off the RG flows before $W$ reaches zero,
and a large but finite conductivity is expected.
In this regime, the system is characterized
by two length scales.  The scattering mean free path $\ell$
is the distance an electron travels in the right moving mode,
say, before suffering an inter-mode backscattering event.
In addition, the thermal length $L_T = v/T$ describes the loss
of phase coherence {\it within} a single mode due to thermal smearing.
On length scales longer than $L_T$, scattering events are uncorrelated\cite{inelastic}.

Following Giamarchi and Schulz\cite{Giamarchi}, the
temperature dependence of $\ell$ may be deduced from 
scaling arguments.  Under a rescaling transformation by a factor ``b"
one can write 
\begin{equation}
\ell(W,T/\omega_c) = b \ell(b^{3-2K}W, bT/\omega_c)  .
\end{equation}
Generally temperature scales as $b^z$, but 
$z=1$ to leading order in $W$.
With the choice $b=\omega_c/T$, the effective temperature
on the right side becomes comparable to the cutoff frequency.
Quantum interference effects should be absent at such high temperatures,
and a perturbative evaluation of the scattering length
should be valid.  Since the scattering rate
should be linear in $W$ one expects $\ell(W,1) = ca/W$,
with some constant $c$.
This gives,
\begin{equation}
\ell(W,T/\omega_c) = 
{{c a} \over {W (T/\omega_c)^{2(K-1)}}}.
\end{equation}

Since $K>3/2$ throughout the conducting phase, $\ell >> L_T$
as $T \rightarrow 0$. 
This implies that successive backscattering events are incoherent,
so that quantum interference effects are absent.  A Boltzmannn
transport description is then appropriate, which relates
the conductivity and scattering lengths: $\sigma = (e^2/h) \ell$, so 
that
\begin{equation}
\sigma \propto (e^2/h)(a/W)(T/\omega_c)^{2(1-K)}.
\end{equation}
In the Appendix we show that this result may also be obtained from the 
Kubo formula, treating the disorder
perturbatively within a Born approximation\cite{Apel}.  
For $1<K<3/2$ it appears naively that perturbation theory should be valid,
since $\ell$ diverges at low temperature.  However, since
$L_T$ diverges faster, successive scattering events become coherent.
This leads to a breakdown of Boltzmann transport and to localization.

To describe the conductivity near the metal-insulator
transition, it is necessary to include the renormalization
of $K$.  The precise temperature dependence of the conductivity in
the crossover region may be determined by integrating both flow
equations (2.14) and (2.15)
out to a temperature
dependent length scale, $\ln b = \ln(\omega_c/T)$. 
Let $\delta \propto K_c - K$ be the ``distance" to the transition,
as shown in Fig. 2.
To leading order in $K - 3/2$, the
renormalized value of $W$ in the
insulating phase, $\delta>0$, is
\begin{equation}
W_R = (8/9) c^2 |\delta|/\sin^2(c\sqrt\delta \ln \omega_c/T).
\end{equation}
The conductivity is then given by (3.5)
with $K = 3/2$ and $W$ replaced by $W_R$,  
\begin{equation}
\sigma \propto (e^2/h) a (\omega_c/T) \sin^2 (c\sqrt{|\delta|} \ln \omega_c/T)
/c^2 |\delta|.
\end{equation}
For the conducting phase $\delta<0$ the expressions are similar, except
``sin" is replaced by ``sinh".  These expressions are accurate provided
$\delta << 1$ and $W_R << 1$.  The latter condition is equivalent
to $L_T << \xi$, where the localization length $\xi$ is given in (3.2).

\begin{figure}
\epsfxsize=3.5in
\centerline{\epsffile{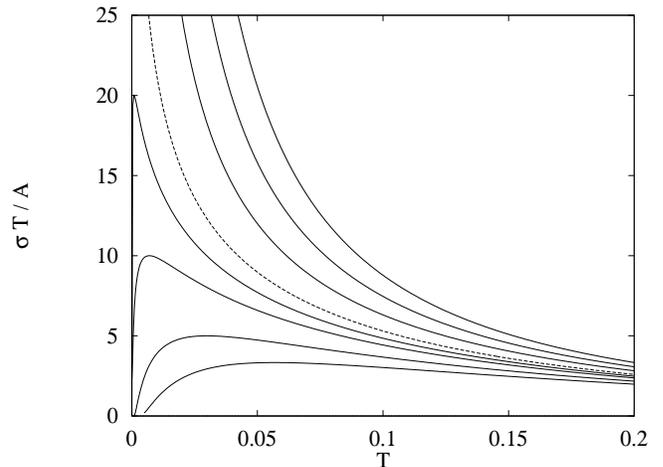}}
\caption{Temperature dependence of the conductivity obtained from (3.7)
for
$c^2 \delta = -.3,-.2,-.1,-.05,0,.1,.2,.3$ ($\sigma$ decreasing).
Negative (positive) values of $\delta$ correspond to the metallic
(insulating) phase.  The dashed line corresponds to $\delta=0$ -
precisely at the metal-insulator transition.
}
\end{figure}

At high temperatures, the conductivity in the transition region is
dominated by the prefactor in (3.7), 
$\sigma =  A/T$, where $A \sim (e^2/h) a \omega_c$.  
In Fig. 3 we plot the temperature
dependence of $\sigma T/A$ from (3.7) for several values of $\delta$ above and below
the transition.  On the metallic side, $T\sigma/A$
diverges as $(T/\omega_c)^{-\alpha}$ at low temperatures, where the exponent $\alpha = c\sqrt{\delta}$.
As the transition is approached from above, $\alpha \rightarrow 0$,
and logarithmic corrections to the power law behavior develop.  Precisely
at the transition, the conductivity varies as
\begin{equation}
\sigma T/A = \ln^2 \omega_c/T.
\end{equation}
Slightly below the transition $\sigma T$ increases logarithmically as the temperature
is lowered, down to a temperature of order $T^* \approx \omega_c \exp [- c/\sqrt{\delta}]$.
Below $T^*$ the conductivity decreases, signaling the crossover to the insulating
regime.  
As the temperature is lowered further $W$ flows out of the perturbative regime
when $\sigma T/A < 1$.  
The conductivity should then follow the Mott variable range hopping law (3.1) with $T_0 \sim T^*$.

\section{Transport with leads}

In the previous section we discussed the temperature
dependence of the line junction conductivity for an
infinitely long wire. 
In practice, of course, the line junction will have some finite
length, $L$.  Moreover, it is initially unclear
how the bulk line junction conductivity can be extracted from
a standard Hall conductance measurement.
In this section we discuss how this can be achieved.

Consider the geometry shown in the Fig. 1.  A Hall bar of width $L$
is cut into two by a line junction.  
The easiest quantity to measure experimentally is the Hall conductance,
passing a current from source to drain, and measuring the voltage
drop between the two Hall voltage probes (denoted 1 and 2)
which straddle the line junction.  Ignoring contact resistances
at the source and drain electrodes, this is equivalent to the
two-terminal source-to-drain conductance,
\begin{equation}
G_{sd} = I_{sd}/(V_s - V_d)  ,
\end{equation} 
where $V_s$ and $V_d$ are the source and drain voltages.

Imagine starting in equilibrium with $V_s = V_d =0$,
and then raising the source voltage to $V_s = V$.
This injects an extra incident current, $I_{in} = (\nu e^2/h)V$,
from the source electrode along the top edge.  At the line junction
this current splits, with a current $I$ passing along the line junction,
and $I_{in} -I$ continuing along the edge into the drain electrode.
The transmitted source-to-drain current is thus
$I_{sd} = I_{in}-I$.  Since
the two ends of the line junction
are separated by the voltage $V$,
the current flowing in the line junction
is $I=GV$, where $G$ is the two terminal {\it conductance}
of the line junction.
The source-to-drain conductance is thereby related to the
line junction conductance;
\begin{equation}
G_{sd} = \nu e^2/h - G  .
\end{equation}

The two terminal conductance of the line junction depends on the length $L$ of the junction.
Provided $L$ is long compared to the thermal coherence length $L_T$, a classical
description is possible in terms of the bulk {\it conductivity} of an infinitely
long line junction.  In particular, solving a Boltzmannn equation\cite{KF3} subject to
the boundary condition that the channels incident to the left and right of the
line junction are separated by a voltage $V$, one arrives at the two terminal
conductance
\begin{equation}
G = \nu {e^2\over h} {\sigma \over {\sigma + \nu (e^2/h)L}}.
\end{equation}

In the localized phase,
$\sigma << (e^2/h) L$, so that (4.3) reduces to the classical expression $G = \sigma/L$, with a small conductance.
In the extended phase this may also be true at high temperatures.  However,
upon cooling $\sigma$ grows, and the backscattering length
$l$ will eventually
become comparable or larger than the line junction length $L$.
In this opposite limit, $\sigma >> (e^2/h) L$, an electron
will typically be transported all the way along the line junction length
without suffering any backscattering collisions.
The line junction conductance will be very close
to perfect, $G \approx \nu e^2/h$, whereas the source-to-drain
conductance will be much smaller than the quantum unit $G_{sd} << \nu e^2/h$.
In this low temperature regime of the metallic phase, one thus has
$G_{sd} \propto (e^2/h) (L/a)W(T/\omega_c)^{2(K-1)}$.

Finally, at very low temperatures
in the metallic phase, when $L_T>>L$, the system length $L$ 
replaces temperature as a cutoff.
In this limit the line junction
behaves effectively as a point contact, with ideal leads.
One expects the source-to-drain conductance to vary
as $G_{sd} \propto (e^2/h) (L/a) W(T/\omega_c)^{(2/\nu -2)}$.

To extract the bulk line junction conductivity, and avoid the
complications associated with the various different regimes,
it is clearly desirable to make the line junction as long as possible.
The temperature range should then be restricted so that
the source-to-drain conductance is close to its quantized value.
The bulk conductivity can be readily extracted: 
\begin{equation}
\sigma =  L(\nu e^2/h - G_{sd}){\nu e^2/h\over G_{sd}}.
\end{equation}

\section{localization transition for 2/3}

Hierarchical FQHE states at filling factors different than
$1/\nu$ an odd integer, are believed to have
multiple propagating modes on a single edge.  This necessarily complicates
the analysis of a line junction in such states.  For simplicity, we
discuss only the experimentally most robust hierarchical state - $\nu=2/3$.

We first briefly review the theory of a single edge of a $\nu=2/3$ fluid.
MacDonald and Wen\cite{MacD} originally argued that the edge consists of two modes:
a forward propagating mode similar to a $\nu=1$ edge,
and a backward propagating mode similar to a $\nu =1/3$ edge.
The appropriate Hamiltonian is,
\begin{equation}
{\cal H}_0 = \pi v_1 n_1^2 + 3\pi v_2 n_2^2 ,
\end{equation}
where $n_1 = \partial_x \phi_1/2\pi$ is the charge density
propagating
downstream at velocity $v_1$, and $n_2 = - \partial_x \phi_2/2\pi$
the density propagating upstream at $v_2$.  
These densities satisfy the commutation relations:
\begin{equation}
[n_1(x),n_1(x^\prime)] = (i /2 \pi) 
\partial_x \delta(x-x^\prime) ,
\end{equation}
\begin{equation}
[n_2(x),n_2(x^\prime)] = - (1/3)(i /2 \pi ) 
\partial_x \delta(x-x^\prime) .
\end{equation}
Generally, these two modes will interact via a term of the
form ${\cal H}_{int} \sim v_{12} n_1 n_2$.

Operators which add charge to the edge take the general form
$O_{n_1,n_2} = e^{i(n_1 \phi_1 + n_2 \phi_2)}$, for integer
$n_i$.  These operators add $n_1$ electrons to channel one,
and $n_2$ 1/3-charged Laughlin quasiparticles to mode two,
creating a total charge:  $Q(n_1,n_2) = e(n_1 + n_2/3)$.

To account for the observed quantized hall conductance at $\nu=2/3$,
it is essential to incorporate processes which can equilibrate
the two edge modes.  The dominant process is $O_{1,-3}$ which transfers
a unit charge from mode two to mode one.  This process
must be mediated by impurities, since the two edge
modes will generally be at different momenta.
In Ref. \onlinecite{KF2} we have analyzed in detail the effects
of such impurity induced tunneling processes.
We find that there are two possible phases, depending on the impurity
strength and the inter-channel Coulomb interaction $v_{12}$.
For a very clean edge with small $v_{12}$ the impurity scattering 
scales to zero at low energies, and charge propagates
in both directions.  But for a dirtier edge, the system has a phase
transition into a disorder dominated phase.  In this phase
the two modes re-structure, forming a charge mode, 
\begin{equation}
n_\rho = n_1 - n_2 = \partial_x \phi_\rho/2\pi,
\end{equation}
propagating
downstream, and a neutral mode,
\begin{equation}
n_\sigma = n_1 - 3n_2 = \partial_x \phi_\sigma/2\pi,
\end{equation}
moving upstream.  (More precisely, the actual neutral mode
which propagates is related to $n_\sigma$ by a spatially
random SU(2) rotation - see Ref. \onlinecite{KF2}.)  
The effective Hamiltonian becomes,
\begin{equation}
{\cal H}_0 = (\pi v_\rho/\nu) n_\rho^2 
+ (\pi v_\sigma /2) n_\sigma^2 .
\end{equation}
In the following we consider
the behavior of a line junction, supposing that the $\nu=2/3$ edges
on either side of of the junction are in this disorder
dominated phase.  For this analysis, we will need the ``local" scaling
dimensions, $\delta$, of the edge operators, defined via
$<O(x,\tau)O(x,0)> \sim \tau^{-2\delta}$.
Using the above definitions, one finds,
\begin{equation}
\delta(n_1,n_2) = {3 \over 4}(n_1 + {n_2 \over 3})^2 + {1 \over 4}(n_1
+ n_2)^2   .
\end{equation}

Consider now a line junction, which
will consist of two charge and two neutral modes,
one above and the other below the junction.
We denote these as $\phi_{\rho,a}$ for the ``top"
and $\phi_{\rho,b}$ for the ``bottom" charge mode - and similarly
for the two neutral modes.
Tunneling processes which transfer charge from the top to
bottom are expressed as products, $O^{a \dagger} O^b$.
For example, the operator $O^{a \dagger}_{1,0}(x) O^b_{1,0}(x)$
tunnels an electron from top to bottom at position $x$ along the junction.
The appropriate term to add to the Hamiltonian is,
\begin{equation}
{\cal H}_{1} = \xi(x) O^{a \dagger}_{1,0}(x) O^b_{1,0}(x) + h.c.  
\end{equation}
where again $\xi(x)$ is a random (complex) tunneling amplitude.

Generally, all tunneling processes which transfer charge from top
to bottom edges in integer units of the electron charge are
allowed.  
Of interest are the most relevant (or least irrelevant) of
such operators, or equivalently those with the smallest
scaling dimensions, $\delta$.  There are three of these, with the
same scaling dimension,
$\delta(1,0)=\delta(2,-3) = \delta(1,-3)=1$.
The first two transfer a charge $e$, whereas zero charge is transferred
for (1,-3).  
A perturbative RG calculation for small disorder $W$,
gives
\begin{equation}
\partial W/ \partial \ell = (3-2\Delta)W,
\end{equation}
where $\Delta = \delta_a + \delta_b$ is the total (local)
scaling dimension of the operator $O^aO^b$.

If we ignore any Coulomb interactions between the modes
on the top and bottom sides of the line junction, then we
can use (5.7) to evaluate the scaling dimension, giving $\Delta =2$.
This implies that {\it all} electron tunneling processes are
irrelevant.  The line junction is in a conducting phase, with
the electron backscattering strength scaling to zero at low
energies.  This should be the case when the gate potential is
adjusted so that the top and bottom modes are well separated.

But as the gate potential is reduced, the modes get closer together,
and the Coulomb interaction increases.  Since the neutral modes
do not carry any charge, the Coulomb interaction acts only between
the top and bottom charge modes.  Since these modes move in opposite
directions, a Coulomb interaction will modify the scaling
dimensions, just as in Section 2.  To see this consider
the total Hamiltonian for the clean line junction,
${\cal H}_0 ={\cal H}^a_0 + {\cal H}^b_0$.
From (3.6) this factorizes into a sum of the charge and neutral sectors,
${\cal H}_0 ={\cal H}^\rho_0 + {\cal H}^\sigma_0$,
with
\begin{equation}
{\cal H}^\rho_0 = (\pi v_\rho/\nu)[n_{\rho,a}^2 + n_{\rho,b}^2 +
\lambda n_{\rho,a}n_{\rho,b} ] ,
\end{equation}
where we have included a Coulomb interaction, with (dimensionless)
strength $\lambda$.  Since the charge densities satisfy,
\begin{equation}
[n_{a/b}(x),n_{a/b}(x^\prime)] = \pm (i\nu /2 \pi ) 
\partial_x \delta(x-x^\prime) ,
\end{equation}
one sees that the charge sector for $\nu=2/3$ is formally identical
to the full theory for $\nu^{-1}$ an odd integer, Eqn. (2.1) and (2.2).
As in Section II, we can diagonalize the Hamiltonian ${\cal H}^\rho_{0}$
by
defining fields, $\phi_{\rho,a/b} = \sqrt{\pi}(\phi \pm \theta)$.
In this way, ${\cal H}^\rho_{0}$ takes the form of (2.10) with
a charge stiffness,
\begin{equation}
K_\rho = {1 \over \nu} [{{1- \lambda} \over {1+ \lambda}}]^{1/2} .
\end{equation}

Once the Hamiltonian is diagonalized, one can readily obtain the scaling
dimension for the electron tunneling operator that appears in (5.9),
giving
\begin{equation}
\Delta = {1 \over 2} + {{3 K_\rho } \over 2}  ,
\end{equation}
where the first term is from the neutral sector.
In the absence of Coulomb interactions between the top and bottom charge
modes, $K_\rho =1$, and the tunneling is irrelevant, as before.
However, with increasing Coulomb interaction, $K_\rho$ decreases,
and eventually $\Delta < 3/2$.   When this happens, even weak back
scattering grows under the RG transformation,
and the system scales into a disorder dominated phase.
Although the properties of this phase are perturbatively
inaccessible, it is natural to presume that both the
charge and neutral excitations are localized in this phase.

We thereby conclude that a line junction in a $\nu=2/3$ fluid
should be qualitatively similar to a $\nu=1/3$ junction.
Two phases should be present, a conducting phase
when the modes are well separated, and a localized phase.
Upon tuning the gate potential, one should be able to pass
through the localization phase transition separating the two
phases.  In the conducting phase, the electrical conductivity
should diverge as a power law of temperature, precisely as
in the analysis of Section III.

\section{Conclusion}
A line junction in the fractional quantum Hall effect offers a
unique 
opportunity for observing a one-dimensional localization transition.
In this novel geometry, the 1d system is formally
equivalent to a 1d electron gas with {\it attractive}
interactions.  With sufficiently strong attraction,
localization ceases to be operative in 1d, and the system
can undergo a metal-insulator transition.
The key signature of this 1d metal-insulator transition
is the temperature dependence of the conductivity,
which diverges as a (variable) power of temperature
in the metallic phase.  In the insulating phase
the conductivity is expected to drop rapidly with cooling,
following a variable range hopping law, whereas  
right at the transition a 
$1/T$ behavior with logarithmic corrections is predicted.
As discussed in detail, the 1d conductivity can be extracted
from a standard Hall transport measurement, in the regime where the
Hall conductance is close to it's quantized value.

We are grateful to Sora Cho for helpful conversations.  M.P.A.F.
acknowledges support from the National Science Foundation under grants No. PHY94-07194,
DMR-9400142 and DMR-9528578.  C.L.K. has been supported by grant DMR-9505425.
\appendix
\section{Linear Response Theory for the Conductivity}

Here we evaluate the conductivity in linear response theory,
treating the disorder perturbatively in the Born approximation.
This is valid in the conducting phase, where $W$ scales to zero
at low temperatures and scattering events are uncorrelated.
The Kubo formula expresses the conductivity
as a current-current correlation function:
\begin{equation}
\sigma = {1 \over {\omega_n}}\int_{x,\tau} 
{\cal D}_I (x,\tau) e^{-i\omega_n\tau} |_{\omega_n \rightarrow
i \omega + \epsilon}  
\end{equation}
where
\begin{equation}
{\cal D}_I (x-x',\tau - \tau') = [<T_\tau I(x,\tau) I(x',\tau')>]_{\rm ens}  .
\end{equation}
Here the square brackets denote an ensemble average over realizations of the disorder.  After ensemble averaging the current correlation function
is translationally invariant.  Fourier transforming 
gives,
\begin{equation}
\sigma = {1 \over { \omega_n}} {\cal D}_I(k=0, \omega_n) |_{\omega_n \rightarrow
i \omega + \epsilon}
\end{equation} 

An expression for the current operator can be
obtained by
noting that the total  
one-dimensional density is,
$n = n_R + n_L = (\nu/\sqrt{\pi}) \partial_x \theta$.
Thus $\theta(x)$ is proportional to the integrated charge less than $x$,
so that current operator is,
\begin{equation}
I = {\nu \over {\sqrt{\pi}}} \partial_t \theta  .
\end{equation}
The conductivity then becomes
\begin{equation}
\sigma =  {{\omega_n \nu^2} \over \pi}
{\cal D}_\theta(k=0,\omega_n) |_{\omega_n \rightarrow
i \omega + \epsilon} ,
\end{equation}
where we have defined
\begin{equation}
{\cal D}_\theta (x-x',\tau - \tau') = [<T_\tau \theta(x,\tau) \theta(x',\tau')>]_{\rm ens}  .
\end{equation}
In the absence of any disorder, 
\begin{equation}
{\cal D}_{\theta,0} = { {vK} \over {v^2 k^2 + \omega_n^2}}  .
\end{equation}
With disorder present we write, 
\begin{equation}
{\cal D}_\theta^{-1} = {\cal D}_{\theta,0}^{-1} + \Sigma  ,
\end{equation}
with a self-energy $\Sigma(k,\omega)$.  This self-energy
can readily be evaluated to leading order in the disorder strength.
The relevant diagrams are shown in Fig. 4 and may be evaluated as
\begin{equation}
\Sigma(k,\omega_n) = \Delta_W \int_0^\beta d\tau (1-e^{i\omega_n\tau}) 
\left[ {\pi T / \omega_c \over {{\rm sinh} \pi T \tau}}\right]^{2K}.
\end{equation}
Analytically continuing to real time, this becomes
$\Sigma(k,\omega) = c_{K} i\omega W/a (T/\omega_c)^{2(K-1)}$,
where $c_K$ is a numerical factor which depends on $K$.
Then using (3.10) and (3.13), we recover the conductance given in (3.5).
\begin{figure}
\epsfxsize=3in
\centerline{\epsffile{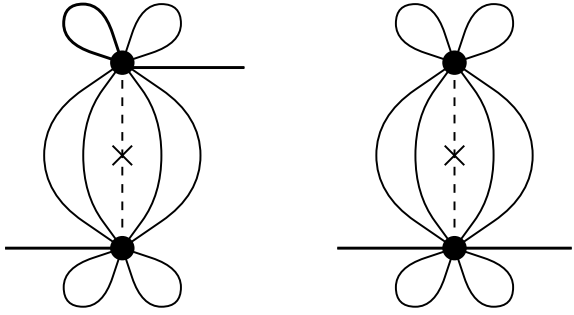}}
\caption{Diagrams for the self energy. The solid lines represent
the propagators ${\cal D}_{\theta,0}$ and the dashed lines
represent the impurity scattering vertex.  A sum over all possible combinations
of these lines is implied.}
\end{figure}

\end{multicols}


\begin{references} 

\bibitem{Wen}
X.G. Wen, Phys. Rev. B {\bf 43}, 11025 (1991); Phys. Rev. Lett.
{\bf 64}, 2206 (1990).  X.G. Wen, Phys. Rev. B {\bf 44} 5708 (1991).
\bibitem{KF1}
C.L. Kane and M.P.A. Fisher, Phys. Rev. B{\bf 46}, 15233 (1992).
\bibitem{Webb}
F.P. Milliken, C.P. Umbach and R.A. Webb, Solid State Commun. {\bf 97}, 309 (1996).
\bibitem{Chang}
A.M. Chang, L.N. Pfeiffer and K.W. West, Phys. Rev. Lett., to appear.
\bibitem{Renn}
S. Renn and D.P. Arovas, Phys. Rev. B {\bf 51}, 16832 (1995).
\bibitem{Giamarchi}
T. Giamarchi and H.J. Schulz  Phys. Rev. B {\bf 37}, 325 (1988).
\bibitem{Schulzreview} 
H. J. Schulz, Int. J. Mod. Phys {\bf 5}, 57 (1991).
\bibitem{Apel}
W. Apel and T.M. Rice, Phys. Rev. B {\bf 26}, 7063 (1982); 
\bibitem{KF2}
C. L. Kane, M.P.A. Fisher and J. Polchinski,
Phys. Rev. Lett. {\bf 72}, 4129 (1994).
\bibitem{inelastic}
For non interacting electrons ($K=1$), coherence between scattering events 
decays with the inelastic scattering length, $L_{\rm in}$ rather than
$L_T$.  However, with interactions ($K\ne 1$), $L_{\rm in} = c L_T$ 
with a coefficient $c$ which diverges when $K\rightarrow 1$.
See, for example, W. Apel and T.M. Rice, J. Phys. C {\bf 16} L271 (1983).
\bibitem{KF3}
C.L. Kane and M.P.A. Fisher, Phys. Rev. B {\bf 52}, 17393 (1995).
\bibitem{MacD}
A. H. MacDonald, Phys. Rev. Lett. {\bf 64}, 222
(1990); 


\end{references}
\end{document}